\documentclass[12pt]{article}
\newcommand{\be}{\begin{equation}}
\newcommand{\ee}{\end{equation}}
\newcommand{\lll}{\langle}
\newcommand{\rrr}{\rangle}
\newcommand{\llll}{\langle\langle}
\newcommand{\rrrr}{\rangle\rangle}
\newcommand{\T}{\mbox{Tr}}
\begin{document}
\title{ Casimir scaling as a test of QCD vacuum. }
\author{V.I.~Shevchenko, Yu.A.~Simonov }
\maketitle
\begin{abstract}
Recent accurate measurements \cite{bali1} of static potentials
between sour\-ces in various representations of
the gauge group $SU(3)$ provide a crucial test of the QCD
vacuum models and different approaches to confinement.
The Casimir scaling of the potential
observed for all measured distances
implies strong suppression of higher cumulant
contributions.
The consequences for the instanton vacuum model
and the spectrum of the QCD string are also discussed.
\end{abstract}

\section{Introduction}

The structure of QCD in its nonperturbative domain
commands attention of
theorists for many years.
Confinement and chiral symmetry breaking have been studied both using
theoretical models and lattice simulations (for review see \cite{sim1}).

 However most of the models
are designed to describe confinement of colour
charge and anticharge in the fundamental representation
of the gauge group $SU(3)$, i.e. the area law for the
simplest Wilson loop and hence linear potential
between static quark and antiquark.
A supplementary and as we shall see below, very important information
about QCD vacuum is provided by the investigation of interaction
between static charges in higher SU(3) representations. Comparing static
potentials for different charges one can derive information about field
correlators in the vacuum , which is not possible to obtain from
fundamental charges alone.

The recent accurate measurements of the corresponding potential
have been performed by G.Bali in \cite{bali1}. Preliminary
physical analysis of the data from \cite{bali1} was presented in
\cite{simhep}. We
present in this paper more extended investigation
and discuss new important information about the QCD
vacuum and constraints on several QCD vacuum models.

\section{Casimir scaling of the static potential}

We define static potential between sources at the distance $R$
in the given
representation $D$ as:
\be
V_D(R)=-\lim_{T\to\infty}\frac{1}{T} \ln \lll W(C)\rrr,
\label{eq1}
\ee
where the Wilson loop $W(C)$ for the rectangular contour $C=R\times T$ in
the "34" plane admits the
following expansion \cite{sim2}
$$
\lll W(C)\rrr = \left\lll {\T}_D\>{\mbox P}\exp\left(ig\int\limits_C
A_{\mu}^a T^a dz_{\mu}\right)\;\right\rrr =
$$
\be
= \T_D \exp \>\sum\limits_{n=2}^{\infty}\int\limits_S (ig)^n \llll F(1)
F(2).. F(n) \rrrr d\sigma (1)... d\sigma (n)
\label{eq2}
\ee
Nonabelian Stokes theorem \cite{nast}
has been used in the above expression with the notation
$F(k)d\sigma(k) =
\Phi(x_0,u^{(k)})E_3^a(u^{(k)})T^a\Phi(u^{(k)},x_0)
d \sigma_{34}(u^{(k)})$,
where $\Phi$ is a parallel
transporter and
$x_0$ is an arbitrary point on the surface $S$
bound by the contour
$C$.
The double brackets $\llll ... \rrrr$
denote irreducible Green's functions
proportional to the unit matrix in the colour
space.\footnote{It makes unnecessary to write colour ordering
operator in
the r.h.s. of (\ref{eq2}). Notice also, that the averages
in (\ref{eq2}) refer to a single Wilson loop and do not take into
account screening effects which add to W(C) multiloop contributions,
as will be explained below in section 4.}

Since (\ref{eq2}) is gauge-invariant, it is convenient
to make use of generalized contour gauge \cite{gcg}, which
is defined by the condition $\Phi(x_0, u^{(k)})\equiv 1$.

The $SU(3)$ representations $D=3,8,6,15a, 10,27, 24, 15s$
are characterized by $3^2-1=8$ hermitian generators $T^a$ which
satisfy the commutation relations $[T^a, T^b] = i f^{abc} T^c$.
One of the main characteristics of the representation
is an eigenvalue of quadratic Casimir operator ${\cal C}^{(2)}_D $,
which is defined according to ${\cal C}^{(2)}_D = T^a T^a = C_D\cdot \hat1$.
Following the notations from \cite{bali1} we
introduce the Casimir ratio $d_D = C_D / C_F$, where
the fundamental Casimir $C_F = (N_c^2-1)/2N_c$ equals to
$4/3$ for $SU(3)$. The invariant trace is given by $\T_D \hat1 =1$.

Since a simple algebra of the rank $k$ has exactly $k$ primitive
Casimir--Racah operators \cite{cr}
of order $m_1,..,m_k$, it is possible to
express
those of higher order in terms of the primitive ones.
In the case of $SU(3)$ the primitive Casimir operators
are given by
\be
{\cal C}^{(2)}_D = {\delta}_{ab} T^a T^b \;\; ; \;\;
{\cal C}^{(3)}_D = {d}_{abc} T^a T^b T^c
\label{eq67}
\ee
while the higher rank Casimir operators are defined as follows
\be
{\cal C}^{(r)}_D = d^{(r)}_{(i_1 .. i_r)} T^{i_1} .. T^{i_r}
\ee
where the totally symmetric tensor $d^{(r)}_{(i_1 .. i_r)}$
on the $SU(N_c)$ is expressed in terms of $\delta_{ik}$ and
$d_{ijk}$ (see, for example,
\cite{hysp}).

The potential (\ref{eq1}) with the definition (\ref{eq2}) admits the
following decomposition
 \be
 V_D(R)= d_D V^{(2)}(R) + d^2_D V^{(4)}(R)+...,
 \label{eq3}
 \ee
where the part denoted by dots contains terms, proportional to
the higher powers
of the quadratic Casimir as well as to higher Casimirs.

The fundamental static potential contains perturbative
Coulomb part,  confining linear and constant
terms
\be
V_D(R) = \sigma_D R - v_D - \frac{e_D}{R}
\label{pppo}
\ee
 The  Coulomb part is now known up to two loops \cite{psr}
and is proportional to $C_D$.
The "Casimir scaling hypothesis"
\cite{go} declares, that
the confinement potential is also proportional to the
first power of the quadratic Casimir $C_D$, i.e.
all terms in the r.h.s. of (\ref{eq3}) are much smaller than
the first one. In particular, for the string tensions
one should get $\sigma_D/\sigma_F = d_D$.

This scaling law is in perfect agreement with the results
found in \cite{bali1}. Earlier lattice calculations
of static potential between sources in higher representations
\cite{go}
are in general agreement
with \cite{bali1}.

To see, why this result (to be more
precise -- why the impressive {\it accuracy} of the "Casimir scaling"
behaviour) is nontrivial,
let us examine the colour structure of a few lowest
averages in the expansion (\ref{eq2}).

The first nontrivial Gaussian cumulant in
(\ref{eq2}) is expressed through
$C_D$ and representation--independent averages as
\be
\T_D\lll F(1) F(2)\rrr =
\frac{C_D}{N_c^2 -1}\> \lll F^a(1)
F^a(2)\rrr =
\frac{d_D}{2N_c}\> \lll F^a(1)
F^a(2)\rrr,
\label{eq4}
\ee
so Gaussian approximation satisfies "Casimir scaling law"
exactly. It is worth being mentioned, that this fact does not depend
on the actual profile of the potential.
It could happen, that the linear potential observed
in \cite{bali1} is just some kind of
intermediate distance characteristics and changes the profile
at larger $R$ (as it actually should happen
in the quenched case for the
representation of zero triality due to the screening of
the static sources by dynamical gluons from the vacuum, or,
in other words, due to gluelumps formation).
The coordinate dependence
of the potential
is not directly related to the Casimir scaling,
 and can be analized
at the distances which are small enough
to be affected by the screening effects.\footnote{The same
is true for the general criticism \cite{diak3}
of the confining potential evaluation on the lattice. The
considered potential might be even different from linear but
still demonstrate Casimir scaling.}

Having made these general statements, let us come back to
our analysis of the contributions to the potential
from different field correlators.
We turn to the quartic correlator and
write below several possible colour structures
for it. We introduce the following abbreviation
$$
\lll F^{[4]} \rrr = \lll F^a(1) T^a  F^b(2) T^b F^c(3) T^c
F^d(4) T^d \rrr
= T^a T^b T^c T^d {\lll F^{[4]} \rrr}^{abcd}
$$
where the Lorentz indices and coordinate dependence
are omitted for simplicity of notation.
One then gets, with some work in the last case
the following possible structures
$$
\begin{array}{ll}
{\lll F^{[4]}\rrr}^{abcd} \sim
{\delta}_{ab}\delta_{cd}&
\lll F^{[4]} \rrr \sim C_D^2 \cdot \hat1 \\
{\lll F^{[4]} \rrr}^{abcd} \sim
{\delta}_{ac}\delta_{bd} &
\lll F^{[4]} \rrr \sim \left(C_D^2 -\frac{1}{2} N_c C_D\right)\cdot
\hat1 \\
{\lll F^{[4]} \rrr}^{abcd} \sim
{f}_{ade} f_{cbe}&
\lll F^{[4]} \rrr \sim -\frac14 N_c^2 C_D \cdot \hat1 \\
{\lll F^{[4]} \rrr}^{abcd} \sim
{f}_{ace} f_{bde} &
\lll F^{[4]} \rrr = 0 \\
{\lll F^{[4]} \rrr}^{abcd} \sim
{f}_{apm} f_{bpn} f_{dem} f_{cen}\;\;\;\;\; &
\lll F^{[4]} \rrr \sim \frac94 \left(
C_D^2 + \frac12 C_D
\right) \cdot \hat1 \\
\end{array}
$$

The operator ${\cal C}^{(3)}_D$ enters together with ${\cal C}^{(2)}_D$
at higher orders.
Notice, that the terms, proportional to the
square of $C_D$ appear in both the $\delta \delta$ parts
(the first and the second strings)
and higher order interaction parts (the last string).
Mnemonically the $C_D$ -- proportional components
arise from the diagrams where the noncompensated
colour flows inside
while the $C_D^2$ -- components describe the interaction
of two white objects. It is seen, that the Casimir
scaling does not mean "quasifree gluons",
instead it means roughly speaking "quasifree white multipoles"
(see discussion at the end of the paper).

Let us analyse the data from \cite{bali1} quantitatively.
We have already mentioned, that the Coulomb potential
between static sources
is proportional to $C_D$ up to the second loop (and possibly
to all orders, this point calls for further study) and hence
we expect contributions proportional to
$C^2_D\sim d^2_D$ to the constant
  and linear terms, i.e. we rewrite (\ref{eq4})
as follows
  \be
  V_D(R)= d_D V^{(2)}(R)+ d^2_D (v_D^{(4)} +  \sigma_D^{(4)} R).
  \label{eq5}
  \ee
and all higher contributions are omitted.
  Here
   $ v_D^{(4)}, \sigma_D^{(4)}$
   measure the $d_D^2$--contri\-bu\-tion of the
  cumulants higher than Gaussian to the constant term and string
  tension respectively.
The results of the fitting of the data from \cite{bali1} with
(\ref{eq5}) for some representations are shown in the Table 1.
See also Fig.1, where the quantity $(V_D(R)- d_D V_F(R))$ versus
distance $R$ is depicted. This figure shows the same data as
the Figure 2 from the paper \cite{bali1}.

\begin{table}
\caption{ {The Casimir--scaling and Casimir--violating string
tensions
and shifts.
Based on the data from G.Bali, hep-lat/9908021.}}
\bigskip
\bigskip

\begin{tabular}{|c|c|c|c|c|c|c|}
\hline
& & & & & & \\
$D$ & $ {\sigma}_D^{(4)}\cdot 10^{4}$ & $\Delta
{\sigma}_D^{(4)}\cdot 10^{4}$
& $v_D^{(4)}\cdot 10^{4}$ & $\Delta v_D^{(4)}\cdot 10^{4}$ & $
\left| \sigma_D^{(4)}/ \sigma_D^{(2)}\right| $& $ \chi^2 / N $
\\
& & & & & &\\
\hline
& & & & & &\\
8 &  -3.486  &   1.2 & -2.513 & 2.8 & 0.004 & 19.22 / 43
\\
& & & & & &\\
\hline
& & & & & &\\
  6 & -6.428 &  1.2 & 0.950  &  2.6 &0.007 & 25.76 / 42
\\
& & & & & &\\
\hline
& & & & & &\\
15a   &-5.244  & 0.55   &-0.5611  &1.1 &0.003   &39.06 /
42 \\
& & & & & &\\
\hline
   &  &   &  & &  &  \\
10 &-4.925 &  0.50 & 0.2489 & 1.0 &0.003 & 22.05 / 41\\
& & & & & &\\
\hline
\end{tabular}
\label{tab2}
\end{table}
All numbers in the table 1 are dimensionless
and given in lattice
units.
The author of \cite{bali1} used anisotropic lattice with the spatial
unit $a_s^{-1} = 2.4 GeV $.

Several comments are in order. First of all it is seen
that the Casimir scaling behaviour holds with very good
accuracy, better than 1\% in all cases in the table 1
with the reasonable $\chi^2$. It should be stressed, that
any possible systematic errors which could be present
in the procedure used in \cite{bali1} must either obey
the Casimir scaling too or be very tiny, otherwise
it would be unnatural to have the matching with such
high precision.\footnote{Since we are mostly interested in the relative
quantities, their actual magnituge in the physical units
is of no prime importance for us. This is another reason
why we do not discuss
possible systematical errors of \cite{bali1} and finite
volume effects. }
Nevertheless, the terms violating the scaling are also
clearly seen. While the value of the constant term $v_D^{(4)}$ is
found to be compatible with zero within the error bars,
it is not the case for $\sigma_D^{(4)}$. We have not found
any sharp dependence of $\sigma_D^{(4)}$ on the representation $D$,
which confirms the validity of the expansion (\ref{eq5})
and shows, that the omitted higher terms do not have
significant effect in this case. Notice the negative
sign of the string tension
correction. In euclidean metric it
trivially follows from the fact, that the fourth order
contribution is proportional to $(ig)^4 >0$ while the
Gaussian term is multiplied by $(ig)^2 <0$ for real $g$.

\section{Casimir scaling and instantons}

From perturbation theory it follows, as it was already mentioned
 that Casimir scaling holds up to the $g^6$ terms.
One might suspect therefore that also nonperturbative
configurations when treated exactly, ensure the Casimir scaling.
This is not true however for the models based
on the classical solutions. As an example of the model which
violates scaling we mention here the model of the
dilute instanton gas.

In the simplest $SU(2)$ case the field strength of one instanton
in the regular gauge is given by
\be
gF_{\mu\nu}^a(x,z) = -\frac{{\eta}_{a\mu\nu}\> 4\pi \rho^2}{[(x-z)^2 +
\rho^2]^2}
\label{eq10}
\ee
where  $z_\mu$ is the position of the instanton and $\rho$
is its size, chosing the Wilson plane to be $"12"$ one has
$\eta_{a12} = \delta_{a3}$.
The average over stochastic ensemble of the dilute instanton gas
implies averaging over (global) color rotation of each
instanton $F_{12} \to \Omega^{\dagger} F_{12} \Omega$,
averaging over instanton positions $z_{\mu}^{(k)}$
$k=1..N$, where $N$ is the total number of instantons
and antiinstantons in the volume $V$ and
weighted averaging over instanton sizes $\rho$.
The latter one is assumed to be performed as the last
step of all calculations.
To the lowest
order in density $\left(\frac{N}{V} \rho^4 \right)$
one must consider one instanton  and sum
up over $k, 1\le k \le N$, hence one can write \cite{siminst}
\be
\lll F(1) .. F(2) \rrr = \lll T^3_{\alpha\beta} .. T^3_{\rho\omega}
\rrr_{\Omega} \cdot \frac{N}{V} \int d^4 z F^3(x_1, z) .. F^3(x_n, z)
\label{eq12}
\ee
The lowest order terms read
$$
\lll T_{\alpha\beta}^3 T_{\beta\gamma}^3\rrr_{\Omega}
= {\delta}_{\alpha\gamma} \frac{C_D}{3}\;\;\; ; \;\;\;
\lll T_{\alpha\beta}^3 T_{\beta\gamma}^3
T_{\gamma\delta}^3 T_{\delta\epsilon}^3 \rrr_{\Omega} =
{\delta}_{\alpha\epsilon} \frac{3 C_D^2 - C_D}{15}
$$

Consider now the rectangular Wilson loop $R\times T$
and perform the expansion of the Wilson loop
with respect to the $\Omega$- and $z_{\mu}$ - average
procedures:
\be
\lll W \rrr_{\Omega , z_{\mu}} = \T_D \left\lll {\mbox P} \exp ig
\int\limits_S F_{12}
d\sigma_{12} \right\rrr_{\Omega, z_{\mu}}
= \T_D \exp (-\Lambda_2 +
\Lambda_4 +..)
\label{eq11}
\ee
The terms $\Lambda_2$ and $\Lambda_4$ generate the
following terms in the potential:
\be
V_D(R) = V_D^{(2)}(R) + V_D^{(4)}(R)
\ee
where $V_D^{(2)}$ is proportional to $C_D$ and behaves as $R^{2}$
at small $R$,
namely
$V_D^{(2)}(R) = {\bar \gamma}^{(2)} R^2/{\rho}^3$ where
${\bar \gamma}^{(2)} = \frac{\pi}{32}\gamma^{(2)}$ with
$\gamma^{(2)} = \frac{C_D}{3}\> 16\pi^{3} \> \frac{N}{V}
\rho^4$
For large distances $R\gg \rho$ one has
$V_D^{(2)}(R) = \sigma^{(2)} R$
where
${\sigma}^{(2)} = \gamma^{(2)}/2{\rho}^2$.

Analogously for $V_D^{(4)}(R)$ one has
\be
V_D^{(4)}(R) = -\frac{1}{24}\> \lim\limits_{T\to\infty} \int d^4z
\frac{(4\rho^2)^4}{T}\frac{N}{V}\left(\frac{3C_D^2 - C_D}{15}\right)
J(R,T)^4
\ee
where
\be
J(R,T) = \int\limits_0^T \int\limits_0^R
\frac{dx\> dt }{((x-z_1)^2 + (t-z_4)^2 + z_{\bot}^2 + \rho^2)^2}
\ee
Straightforward calculation gives at large distances
\be
V_D^{(4)}(R) = \sigma_D^{(4)} R \;\; , \;\;\; \sigma_D^{(4)} =
\frac{N}{V}\rho^2 \> \frac{16\pi^4}{9}\> \frac{3C_D^2 -
C_D}{15}
\label{eq17}
\ee
while in the regime $R\ll \rho, T\gg \rho$ one gets
\be
V_D^{(4)} =
- \frac{N}{V}\> \frac{R^4}{\rho} \> \frac{\pi^6}{320}\> \frac{3C_D^2
-C_D}{15}
\label{eq18}
\ee

It is clear from (\ref{eq17}) and (\ref{eq18}), that
linear asymptotics of $V_D^{(4)}$ at large $R$ occures
rather late, for $R\ge 7\rho$.
One concludes, that the Casimir scaling of the potential
is violated by the term $V_D^{(4)}$.
This was interpreted in \cite{simhep} as
the upper bound on the instanton density, for $v_D^{(4)} \approx 10^{-3}
Gev$ it gives $N/V \approx 0.2\> fm^{-4}$, which is much less than
the instanton density typically used in the literature $N/V \sim 1
fm^{-4}$.

The original $SU(3)$ case for the quark--antiquark potential
in the dilute instanton gas approximation
was considered in \cite{diak2}. One has the following expression for the
potential between static sources:
\be
V(R) = 4\pi \frac{N}{V} \int\limits_0^{\infty}
d\rho \nu(\rho) \rho^3 \frac{1}{d(D)} \sum\limits_{J\in D}
(2J+1) F_J(x)\;\; , \;\;\;\; x=\frac{R}{2\rho}
\label{eq20}
\ee
and the function $F_J(x)$ is given by some cumbersome
double integral which can be found in \cite{diak2}.
Here $d(D)\equiv D$ is the dimension of the representation $D$
and sum over $J=0, \frac12 , 1, .. $ goes over all
$SU(2)$ multiplets for
decomposition of the given $SU(3)$ representation with the
corresponding weights. One has $\sum\limits_{J\in D} (2J+1) = d(D)$
and also
$$
d(D) \cdot C_D  = \frac{N_c^2 -1}{3}\> \sum\limits_{J\in D}
J(J+1)(2J+1)
$$
At small $x$ the functions $F_J(x) \sim x^2$, while at large $x$ the
functions
$F_J(x)$ tend to $J$-dependent constant \cite{diak2}.

Numerically one finds at small distances
\be
V(R) = 1.79\cdot\gamma R^2 \cdot \epsilon_D + {\cal O}(R^4)
\ee
where $\gamma = \pi \frac{N}{V} \int_0^{\infty} d\rho \nu(\rho)
\rho $ and numerical coefficients $\epsilon_D$ for $D=3,8,10$ are given
by
$$
\epsilon_3 \; : \; \epsilon_8 \; : \; \epsilon_{10}
\;\; = \;\; 1\; : \; 1.87 \; : \; 3.11
$$
instead of Casimir scaling results
$ 1\; : \; 2.25 \; : \; 4.5 $.

Similar situation takes place
for the large distance asymptotics of the instanton--induced
potentail. It violates Casimir scaling on the level of 20\%
(see \cite{diak2})
and can be exluded by the present analysis at the
level of $10\sigma$.

So one can see the sharp contradiction between the
dilute instanton gas model calculation
for the quark--antiquark
potential and the Casimir scaling of this
potential found on lattice.
This can be understood in one of two ways.
Either instantons are strongly suppressed in the real(hot)
QCD vacuum (as it was observed in \cite{pv})
while they are recovered by the cooling procedure.
Or else instanton
medium is dense and strongly differs from dilute instanton gas, in such a
way that higher cumulant components of such collectivized instantons
are suppressed. Interesting to note, that linear confinement missing
in the dilute gas, is recovered in this case.

\section{Casimir scaling and QCD string}

There is another important consequence of the observed Casimir
scaling. It comes from the analysis of the confinement
potential as being induced by the QCD string.
In this case one has additional contribution to the confining
potential besides the leading linear term, which comes
from the internal dynamics of the string, in particular, from
the transverse worldsheet vibrations. The simplest model in this
respect is the Nambu--Goto string which action is proportional
to the area of the surface bounded by the static sources
worldlines. It modifies the confining potential with respect to the
classical case (nonvibrating string) as
\be
\sigma R \to \sigma R - \frac{\pi}{12} \> \frac{1}{R} + ...
\label{eq7}
\ee
where the term $-\pi/(12\cdot R)$ will be
referred to as the String Vibration (SV) term
\cite{lush}. Despite the Nambu--Goto string
model cannot be rigorously defined in $D=4$, and, in particular
the expansion of the r.h.s. of (\ref{eq7})
meets singularity at the distances $R\sim 1/\sqrt{\sigma} $
it is instructive to look whether or not the data \cite{bali1}
support the existence of such term. It is also worth noting, that
the dimensionless coefficient $-\pi/12$ is determined by the
only two factors: target space dimension
and the chosen string model. Having both factors fixed, it
cannot be freely adjusted.
Assuming $\sigma_D = d_D \sigma_F$, it
is easy to see, that the
Nambu--Goto induced SV term violates Casimir
scaling.

It is a nontrivial task
to separate the contributions of the discussed sort in the
confining potential as it is because these corrections
are essentially large distance effect, where they are
subleading.
But they have to become pronouncing in the expression (\ref{eq5})
due to scaling violation.
Namely one has
\be
V_D(R) - d_D V_F(R) = (d_D -1) \frac{\pi}{12} \frac{1}{R} + ..
\label{eq45}
\ee
where the dots denote the terms, omitted in (\ref{eq7}).
The dashed line on Fig.1 corresponds to the r.h.s. of (\ref{eq45}).
It is seen, that Casimir scaling-violating
SV  term of the form (\ref{eq7}) is completely
excluded for all measured distances.\footnote{
 Notice,
that even the
sign of the leading Nambu--Goto string correction is opposite
to what has actually been observed.}

We should mention at this point, that
the lattice measurements performed in \cite{green} did also
demonstrate no fingerprints of the SV corrections
of the form discussed.

One can see different explanations of this result.
First of all, nobody has proved up to now, that the
simplest bosonic Nambu--Goto string model properly
describes the dynamics of the QCD string and
theoretical background of (\ref{eq7}) is not clear.
Just the opposite is true -- there are many resons, why it is
not the case (see discussion in \cite{pol}).
The theory of the QCD string -- whatever it will be --
must explain the observed scaling of the potential.

\section{Screening and string breaking}

In the previous discussion the effect of string breaking in the triality
zero
representation was not taken into account.
The modification of the potential (\ref{pppo})
  due to this effect was considered in \cite{ghp} in the strong coupling
expansion, resulting in the expression for the adjoint Wilson loop  in the
  large $N_c$ approximation
\be
 \lll W_8(C)\rrr = \exp(-\sigma_8\cdot {\mbox{area}}(C)) +
\frac{1}{N_c^{2}}\>
\exp(-4\sigma_3
\cdot {\mbox{perimeter}}(C))
\label{4.1}
\ee
  The appearence of the second term on the r.h.s. of (\ref{4.1})
signals
screening due
to the lattice diagrams related to the vacuum average of the product of two 
undamental
Wilson loops,
while the
  first term is represented by the only one
loop and has the cumulant
expansion (\ref{eq2}).
  We now demonstrate that a similar form also appears in the general
background
  perturbation
theory \cite{simein}, and estimate the corresponding gluelump masses.
  To this end we represent $A_{\mu} = B_{\mu} + a_{\mu}$ and use the
't Hooft identity to average
  separately over $B_{\mu}$ and $a_{\mu}$, while gluon action is written
as
\be
S(B,a)=S_0(B)+S_1+S_2+S_{int}
\label{4.2}
\ee
In this espression $S_1$ is proportional to the first power of $a_{\mu}$,
while for $S_2$ one has
\be
S_2=a^{\mu} {\left( {G[B]}^{-1} \right)}_{\mu\nu} a^{\nu}
\label{4.3}
\ee
and $S_{int} $ contains higher order terms in the field $a_{\mu}$.
We integrate over ${\cal D} a_{\mu}$,
making perturbatve expansion in $S_{int}$, which leads to
\be
\lll W(B+a)\rrr = \int {\cal D} B
\exp(-S(B))\>[{\mbox{Det}}\> G[B] ]^{-\frac12} \; (W(B)+...)
\label{4.4}
\ee
  where dots stand for higher order terms in the expansion over
perturbative fields.
  Note that the retained in (\ref{4.4}) term is purely nonperturbative.
  Using the world-line Fock--Schwinger
representation for $\mbox{Det} G$,
one has
$$
  \mbox{Det}\> G=\exp(\T \ln G)=\exp\left(\>\T \int\limits_0^{\infty}
\frac{ds}{s}
\>\int {\cal D} z_{\mu}\;
\exp(-K)\cdot \right.
$$
\be
\left.
\cdot W_{C_z} (B)\; \exp(\int\limits_C d\tau \> 2g\hat F )\right)
\label{4.5}
\ee
where the integral is to be taken with the standard Wiener measure.
It is seen from (\ref{4.5}) and (\ref{4.4}) that expansion of the exponent
yields expansion in
products of Wilson loops, and the first terms are
 \be
  \lll W_C(B+a)\rrr = \lll W_C(B)\rrr
+ \llll W_C(B)W_{C_z}(B)\rrrr + ..
\label{4.6}
\ee
where the average in the last term include the average over $C_z$
according to (\ref{4.5}).
 The second term in (\ref{4.6}) is responsible for the screening,
since the asymptotics of it for large $C$ is of perimeter type,
rather than area law.
  To find the logarithm of (\ref{4.6}) explicitly, one can use the
Hamiltonian formalism, obtained via the transition
\be
  \int {\cal D} z \;\exp(-\int d^4 x\>  L)= \lll x|\exp(- H T)|y\rrr
\label{4.7}
\ee
  Then for the asymptotics of the second term in (\ref{4.6}) one gets
  \be
  \llll W W \rrrr  = {\mbox{const}} \cdot  \exp(-2 M_{GL}\cdot T)
\label{4.9}
\ee 
For simplified estimates we disregard the last spin term in
 (\ref{4.5}) and interaction of two adjoint gluelumps 
with masses $M_{GL}$.
  To find $M_{GL}$ from $H$ one can use the standard technic of einbein
formalism (see \cite{simein} and references therein)
to get an estimate
\cite{ghp}
\be
M_{GL} \approx 1.4\> GeV
\ee
  This leads to an estimate of
the screening distance $R_0$ from the relation $V_{adj}(R_0)=2
M_{GL}$ to be $R_0\approx 1.4\> Fm$, which is beyond the distance
where Casimir scaling was measured in \cite{bali1}.
It is interesting to note that a similar estimate of gluelump mass
for higher $D$ leads to the decreasing $R_{0}(D)$, e.g. for $D=15s$
one gets $R_0=0.7\> Fm$ which is not in contradiction with the 
data from \cite{bali1}.

\section{Discussion}

The Casimir scaling behaviour of the confining
potential confirmed in \cite{bali1} with the
unprecendented precision leads to many important
consequences, some of which have been discussed
in the present paper.
It is instructive to compare the
pictures of the QCD vacuum, suggested in
different models
from the Casimir scaling
point of view.

Abelian projection language being in wide use nowadays as
one of the most adequate for the dual Meissner scenario
of confinement encounters difficulties in explanation
of the Casimir scaling. The reader is referred to the paper
\cite{poulis} where the question is discussed in details for the
adjoint static charges. The observed adjoint string tension
(at intermediate distances) arises from the interaction
of diagonal abelian projected gluons with the part of the
adjoint source doubly charged with respect to the Cartan
subgroup. If one naively omits the corresponding Faddeev--Popov
determinant it gives $\sigma_{adj} = 4 \sigma_{fund}$.
 It is expected, that the loop expansion of the determinant
produces terms, correcting the above behaviour to the Casimir
scaling relation. Up to the authors' knowledge, it has not yet
been shown analytically, while there are numerical evidences
from the lattice in favour of this possibility (see \cite{poulis}
and references therein). From physical point of view
to reproduce Casimir scaling, which is genuine nonabelian
feature, one needs to restore the original nonabelian
gauge invariance broken by hand in the abelian projected
method.

The now popular  confining mechanism is
the model of fat center vortices \cite{faber}.
While the original center vortex picture
cannot explain confinement of the adjoint charges,
the introduction of the finite thickness of the vortex
makes it possible, to obtain  approximate Casimir ratios
for the string tensions \cite{faber}. However, with the high accuracy
of the data \cite{bali1}
these ratios are exluded.

In the gauge--invariant formalism \cite{sim2}, the Casimir scaling has two
important features. First, it is the direct consequence of the Gaussian
dominance hypothesis (see review
\cite{sim1}) since gaussian correlator
provides the exact Casimir scaling. On the other hand, it implies the
cancellations of
$C_D^2$ -- proportional terms and higher ones in the cluster expansion
(\ref{eq2}). Physically, it means the picture of the vacuum,
made of relatively small colour dipoles with weak interactions
between them.
One can imagine two possible scenario.
According to the first one, Casimir scaling is the consequence of Gaussian
dominance. It this case
any higher cumulant contributes to physical quantities
much less than the gaussian one due to  dynamical reasons.
 There is also the second
possibility, when each higher term in the expansion (\ref{eq2})
is not small, but their sum demonstrates strong cancellations
of Casimir scaling violating terms. These pictures are in close
 correspondence to the stochastic versus coherent vacuum
scenario \cite{sim1}.
This set of questions certanly deserves further study.

There are also several open questions of computational
origin. Additional
measurements are needed in order to clarify the
validity of scaling for higher representations where
the statistics is still rather poor. It would also be very
interesting to establish the adjoint string breaking scale
in $SU(3)$ which could shed some light on the gluelumps physics.

Needless to say, that deeper theoretical understanding
of the QCD vacuum structure is still required.
The ability to incorporate such nontrivial feature as Casimir scaling
is a  necessary property of any reasonable confinement
model.


{\bf Acknowledgements}

The numerical calculations have been done using
the lattice data by G.Bali, the Humboldt University, Berlin. The
authors are very grateful to him for the submitting of
his data and valuable explanations.

The authors thank M.I.Polikarpov and D.I.Diakonov for useful comments.
One of the authors (Yu.S.) is grateful for warm hospitality at the
Institute fo Theoretical Physics, Utrecht and
acknowledges useful discussions with
G.'t Hooft, N.G.Van Kampen and J.A.Tjon.

This work was supported in part by the joint RFFI-DFG grant
96-02-00088G and by the grant RFFI 96-15-96740 for scientific
schools.
V.Sh. acknowledges the support from
ICFPM-INTAS-96-0457.

\newpage

\begin{figure}
\caption{The Casimir scaling results for the adjoint potential,
based on the data from \cite{bali1}: axes -- $(V_{8}(R) - 2.25\cdot 
V_3(R))$;
solid line -- leastsquare fit according to (\ref{eq5}); dashed
line -- scaling violating contribution from the SV term (\ref{eq45}).
All quantities are given in the lattice units.}
\end{figure}

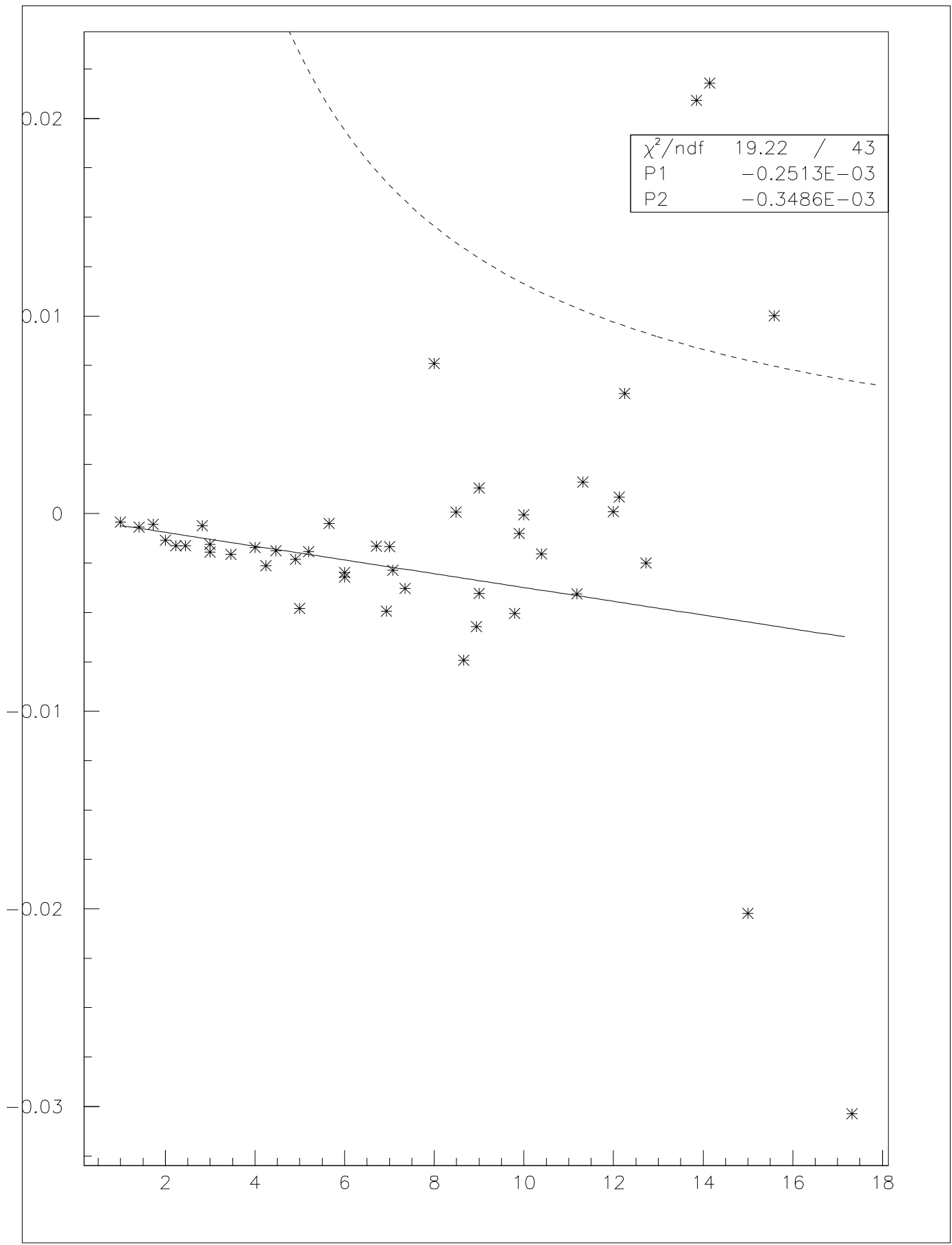
\end{document}